\documentclass[12pt]{article}

\usepackage[english]{babel}

\usepackage[letterpaper,top=2cm,bottom=2cm,left=3cm,right=3cm,marginparwidth=1.75cm]{geometry}

\usepackage{graphicx}
\usepackage{multirow}
\usepackage{vmargin, epsfig, amsmath, amssymb, amsfonts}
\usepackage{amsthm}
\usepackage[title]{appendix}
\usepackage{xcolor}
\usepackage{textcomp}
\usepackage{hyperref}
\hypersetup{colorlinks=true, allcolors=blue}
\usepackage{listings}
\usepackage{booktabs}
\usepackage{authblk}
\usepackage{pdflscape}
\usepackage[numbers, square, compress]{natbib}
\usepackage{xr}
\usepackage{lineno}

\title{Cloud droplet size distribution and optical properties only weakly linked to aerosol size}

\author[1]{Kadja Flore Gali}
\author[1]{Hamed Fahandezh Sadi}
\author[1]{Jesse C. Anderson}
\author[2]{Payton Beeler}
\author[2]{Aaron Wang}
\author[3]{David Richter}
\author[1]{Raymond A Shaw}
\author[4]{Fan Yang}
\author[1,*]{Will Cantrell}
\author[2,*]{Laura Fierce}

\affil[1]{Michigan Technological University}
\affil[2]{Pacific Northwest National Laboratory}
\affil[3]{University of Notre Dame}
\affil[4]{Brookhaven National Laboratory}

\begin{document}

\maketitle

\begin{abstract}
    Changes in aerosol concentrations can modify cloud brightness, producing a strong but poorly constrained influence on Earth’s energy balance. Because cloud reflectivity depends on the size distribution of cloud droplets, and aerosol size strongly governs activation into droplets, one might expect cloud properties to be sensitive to aerosol size distributions. Here we show, through a combination of cloud chamber experiments and high-resolution simulations, that cloud microphysical and optical properties are often insensitive to aerosol size. Detectable impacts on cloud optical properties occur only under weak convective forcing and high aerosol concentrations. These results indicate that, in most conditions, cloud reflectivity can be predicted from aerosol number alone without detailed knowledge of aerosol size distributions, providing new constraints on how aerosol perturbations affect climate.
\end{abstract}

\newpage

\section*{Introduction}
Cloud reflectivity depends on the number and size of droplets within a cloud. Human emissions have increased atmospheric aerosol concentrations, brightening many clouds and producing a major, but uncertain, cooling effect on climate \cite{change2001working,masson2021ipcc,portner2022ipcc,lee2023ipcc}. While the role of aerosol number in altering cloud properties has been extensively studied \cite{TwomeyandWarner,Twomey1974,Twomey1977,Albrecht1989,PincusandBaker,reutter2009aerosol,Li2024GRL}, the contribution of aerosol size distributions remains less clear.

Because aerosol size strongly governs whether particles activate as cloud condensation nuclei (CCN) \cite{Dusek2006}, one might expect cloud droplet size distributions—and hence cloud optical depth—to be highly sensitive to aerosol size. Yet most prior work has explored this question only in simplified frameworks, such as CCN spectra as a function of supersaturation or single-parcel updraft simulations \cite{Hudson2007,hudson2020ccn,reutter2009aerosol,che2017prediction}. At the other extreme, global Earth System Models represent aerosols and clouds at such coarse resolution that potential size effects cannot be mechanistically captured \cite{allwayin2024locally,tang2023earth,fanourgakis2019evaluation}. Direct field observations are likewise limited, since aerosol and cloud properties vary rapidly in space and time, making it nearly impossible to disentangle the role of aerosol size from co-varying factors such as updraft velocity, humidity, and turbulence.

Here we directly probe how aerosol size influences cloud properties using laboratory convective clouds in the Pi Chamber --- a laboratory cloud chamber at Michigan Tech, designed to study aerosol-cloud interactions in turbulent environments through Rayleigh-Bénard convection ---, which captures turbulent conditions more representative of the atmosphere than do parcel models or static CCN measurements \cite{chandrakar2016aerosol,chang2016laboratory,thomas2019scaling,prabhakaran2020role,anderson2023enhancements}. Contrary to our expectations, we find that cloud droplet size distributions and optical proprieties respond only weakly to aerosol size, with detectable effects appearing only under polluted conditions and weak convective forcing. These results demonstrate that, once particles are large enough to activate, cloud reflectivity is controlled primarily by aerosol number rather than size---a mechanistic constraint that helps narrow the uncertainty in aerosol–cloud interactions and their role in Earth’s energy balance.

\section*{Droplet sizes in laboratory-generated clouds only weakly linked to aerosol sizes}\label{sec:response}

To quantify how aerosol size influences the cloud droplet size distribution, we conducted a coordinated set of Pi Chamber experiments \cite{chang2016laboratory} and large-eddy simulations (LES) using the System for Atmospheric Modeling (SAM), adapted to the chamber's geometry and operating conditions \cite{thomas2019scaling,yang2022large,yang2023intercomparison,wang2024designing,wang2024dual,wang2024glaciation,yang2025microphysics}. We performed twelve experiments in which dry NaCl particles (50, 100, and 250 nm) were injected under low and high aerosol loadings (clean vs. polluted) and under strong and weak convective forcing (proxy for updraft strength). Because hygroscopicity and size both modulate the critical supersaturation for activation, a weak response to size implies a similarly weak response to composition. Experimental and numerical details are provided in ~\nameref{sec:Methods} and the Supplementary Information.

\begin{figure*}[htp!]
    \centering
    \includegraphics[width=1.0\linewidth]{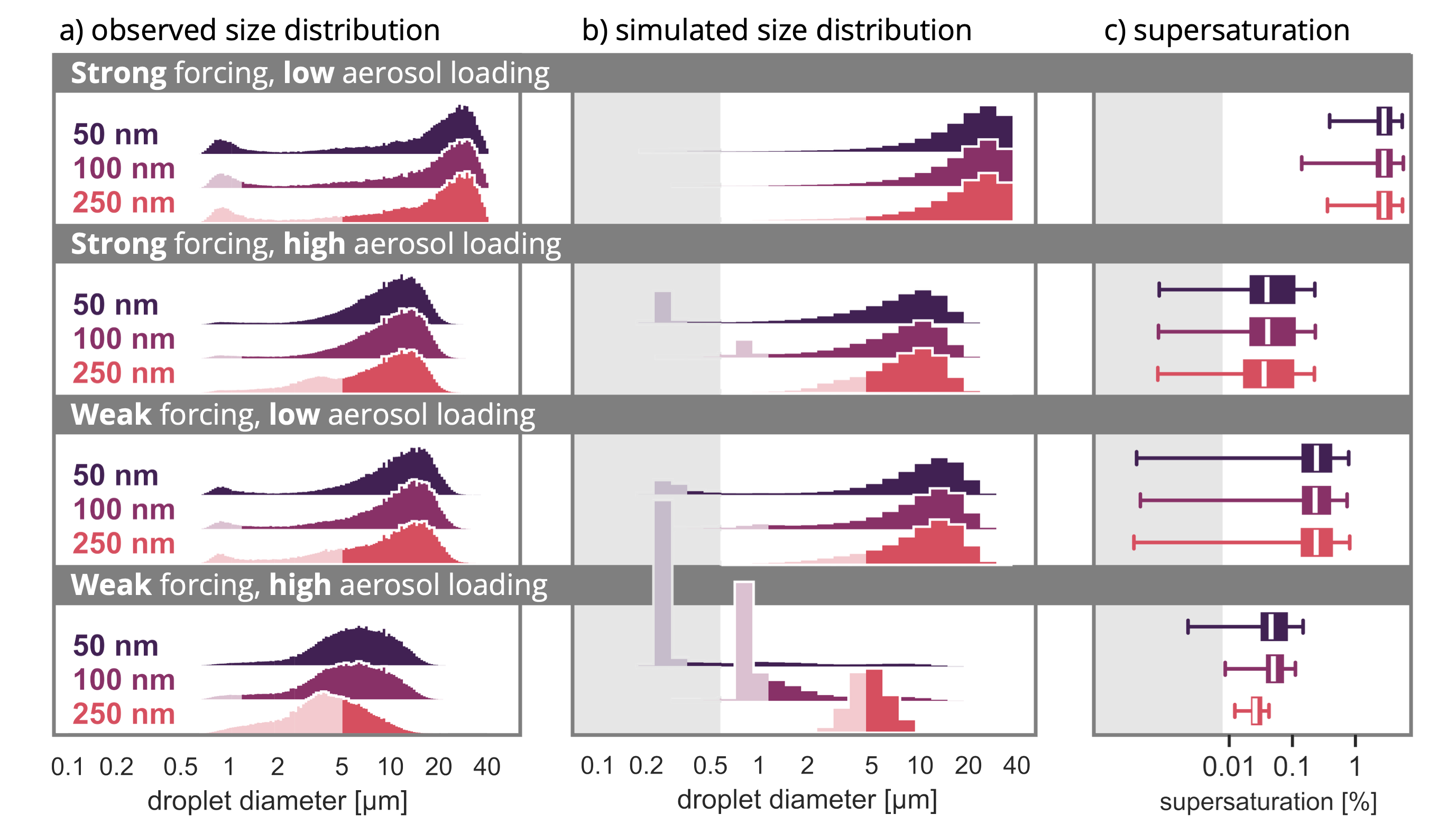}
    \caption{Cloud droplet size distributions and supersaturation fields are generally insensitive to aerosol size, except under polluted, weakly forced conditions. For each experiment (rows), the (a) observed and (b) simulated size distributions within the Pi Chamber domain are only weakly sensitive to the size of injected particles (colors) under most conditions, and the observations and simulations for each experiment are broadly consistent for each experiment (comparison between columns a and b). This weak sensitivity in the droplet size distribution to aerosol size corresponds to a (c) weak sensitivity in the simulated water vFigureapor supersaturation. Only under conditions of weak forcing and high aerosol loadings (bottom panel) did we find significant variation in the size distribution or supersaturation with aerosol size. Under these conditions of weak forcing and high aerosol concentrations, the LES reveals high concentrations of small particles that are below the detection limit of the Welas (shown by grey shading in column  b). Box plots (c) shows median (vertical line), inner-quartile-range (IQR) (boxes), and ranges of 1.5 IQR (whiskers); supersaturation is shown on a symmetric logarithmic log scale, with values below zero shown by the grey shading in column c.}
    \label{fig:size_distribution}
\end{figure*}

Figure~\ref{fig:size_distribution} summarizes results for the 12 cases: each row corresponds to a distinct experimental regime, with columns showing observed droplet size distributions (a), LES output (b), and the simulated distribution of water vapor supersaturation, $s$ (c). With the exception of polluted conditions under weak forcing (bottom panel of Fig.~\ref{fig:size_distribution}), the number and shape of the droplet distribution above 5 $\mu$m show little dependence on injected aerosol size. We emphasize $D>5~\mu$m because these droplets dominate cloud reflectivity. Despite the fact that increasing dry diameter of the aerosol particles from 50 to 250~nm reduces the critical supersaturation for activation by nearly an order of magnitude, the resulting droplet size distributions are nearly indistinguishable across aerosol sizes in most regimes.

Model–observation differences are minor and attributed to limitations in the measurements and SAM's numerical representation. In the SAM LES, the haze-capable microphysics resolves interstitial bins that are not captured by the Welas optical particle counter \cite{yang2023intercomparison}, indicated by the grey shading in Fig.~\ref{fig:size_distribution}b. The apparent absence of these features in Fig.~\ref{fig:size_distribution}a reflects the Welas detection limit (see ~\nameref{sec:Methods} for ranges and SI discussion).

To better understand how the cloud environment modulates the sensitivity of the cloud droplet size distribution to the characteristics of the underlying aerosol particles, we analyzed differences in the water vapor supersaturation, $s$, within the LES across the different cases (Fig.~\ref{fig:size_distribution}c). Within the LES, $s$ is determined within each 6.25~x~6.25~x~6.25~cm$^3$ grid cell volume from the temperature and water vapor mixing ratio tracked by the LES. This simulation data provides details of small-scale variability in $s$ that cannot be accurately measured directly within the Pi Chamber \cite{anderson2021AMT, anderson2024JAS} and is not resolved in large-scale cloud simulations. Fig.~\ref{fig:size_distribution}c shows the mean and variation in $s$ for the three dry sizes that we tested (colors) and cloud conditions (groups). Differences in both the mean and spread of $s$ across regimes (Fig.~\ref{fig:size_distribution}c) align with the observed insensitivity of droplet size to aerosol size.

Under conditions of strong forcing and low aerosol loading (top panel of Fig.~\ref{fig:size_distribution}), where $s$ attains its highest values, the droplet size distributions are effectively independent of injected size. In this regime, nearly all particles activate rapidly and rarely experience subsaturated conditions; solute effects are minimized, and activated droplets behave similarly to pure water, yielding overlapping size distributions across aerosol sizes.

At the other extreme---weak convective forcing and high aerosol loading (bottom panel of Fig.~\ref{fig:size_distribution})---increasing the size of the injected aerosol reduces the concentration of largest cloud droplets and alters the overall cloud droplet size distribution. Either weakening convective forcing in the chamber or increasing aerosol loading lowers the mean $s$ within the chamber (Fig.~\ref{fig:size_distribution}c), which enhances competition for vapor and makes activation more selective. While small differences appear for droplets between 1 and 5 $\mu$m under intermediate regimes, pronounced differences for $D>5~\mu$m arise only under weak forcing and high pollution levels (bottom panel), coincident with the lowest $s$ levels in the chamber.

\section*{Cloud optical properties more responsive to aerosol size under polluted, weakly forced regimes}

\begin{figure*}[htp!]
    \centering
    \includegraphics[width=1.0\linewidth]{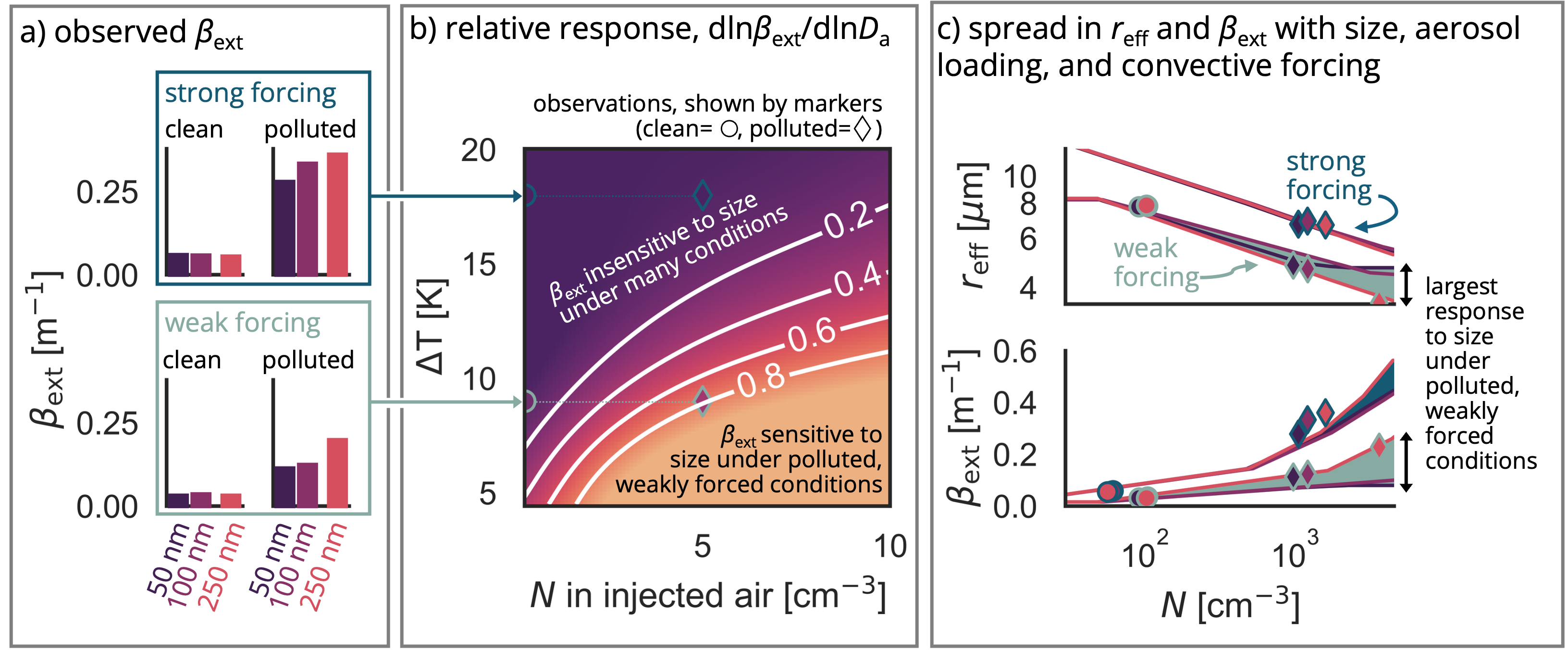}
    \caption{Cloud optical properties show a strong dependence on aerosol size only under polluted, weakly forced regimes, consistent across experiments and LES. The weak response in the droplet size distribution to aerosol size shown in Fig.~\ref{fig:size_distribution} leads to a similarly weak sensitivity in the (a) observed extinction coefficient, $\beta_{\text{ext}}$, except under conditions of weak forcing and high aerosol loadings. To quantify how the environmental regime impacts the response in cloud optical properties to aerosol size, we preformed LES across a wide range of chamber conditions and quantified the relative response in extinction to a relative change in aerosol size, $d\ln\beta_{\text{ext}}/d\ln D_{\text{a}}$, shown in panel (b); predictions were interpolated from the LES output using a Gaussian Process Regressor with the Matern kernel for moderate smoothness. Convective forcing in the Pi Chamber is controlled by the temperature difference ($\Delta T$) between the chamber's floor and ceiling. Observations of $d\ln\beta_{\text{ext}}/d\ln D_{\text{a}}$ agree well with model predictions, as shown by markers in (b), where edge colors indicate strong vs. weak forcing and marker shape indicates different injection rates. The spread in cloud droplet effective radius $r_{\text{eff}}$ and $\beta_{\text{ext}}$ with aerosol size is shown by the shading in panel (c), further illustrating that cloud microphysical responses to aerosol size depend on environmental and pollution regime; model predictions shown by the lines and shading are broadly consistent with observations (markers).}
    \label{fig:extinction}
\end{figure*}

We quantified how the microphysical repsonse to differences in aerosol size translate to changes in cloud optical properties through a numerical investigation of the extinction coefficient, $\beta_{\text{ext}}$, which was computed from the observed and simulated cloud droplet size distributions 
(Fig.~\ref{fig:extinction}). Across both LES and experiments, cloud optical properties are largely insensitive to aerosol size under clean conditions, regardless of forcing: increasing injected dry diameter from 50 to 250~nm yields nearly the same $\beta_{\text{ext}}$ (shown for observations in Fig.~\ref{fig:extinction}a). A strong sensitivity emerges only under polluted, weakly forced conditions, where extinction more than doubles between 50 and 250~nm seed sizes. Under polluted–strong forcing conditions, the Pi Chamber shows a weaker but noticeable size effect ($\sim$30\% higher $\beta_{\text{ext}}$ for 250~nm vs. 50~nm), whereas the LES requires higher concentrations before a comparable response appears (not shown). We attribute this LES–experiment difference to subtle configuration mismatches between the modeled and real chamber and to uncertainty in polluted injection rates.

To systematically characterize regime sensitivity, we analyzed $\beta_{\text{ext}}$ across 45 LES cases and evaluated the relative sensitivity of $\beta_{\text{ext}}$ to differences in the injected aerosol diameter $D_{\text{a}}$. We determined $d\ln\beta_{\text{ext}}/d\ln D_{\text{a}}$ for each combination of convective forcing and pollution conditions (see Fig.~\ref{fig:extinction}b). The sensitivity to aerosol size increases with aerosol loading at fixed forcing, and the loading threshold for a noticeable response depends on the strength of convective forcing. For example, under weak forcing (e.g., $\Delta T=9$ K), $d\ln\beta_{\text{ext}}/d\ln D_{\text{a}} \gtrsim 0.1$ for injection rates $\gtrsim 0.25$ cm$^{-3}$ s$^{-1}$, implying that a factor-of-two increase in $D_{\text{a}}$ increases $\beta_{\text{ext}}$ by $\gtrsim$7–10\%. Under strong forcing (e.g., $\Delta T=19$ K), injection rates must exceed $\sim$8 cm$^{-3}$ s$^{-1}$ to reach a similar elasticity. This regime dependence is consistent between LES (surface) and observations (markers), with minor quantitative offsets attributable to configuration and injection-rate uncertainties noted above.


These regime dependencies are further illustrated in Fig.~\ref{fig:extinction}c. For all forcing regimes and aerosol sizes, cloud effective radius ($r_\text{eff}$) decreases with increasing aerosol loading (top row panel of Fig.~\ref{fig:extinction}c), reflecting stronger competition for water vapor among more numerous droplets. Whereas $r_{\text{eff}}$ shows little variation with size under strong forcing (narrow spread dark blue shading), the sensitivity of $r_\text{eff}$ to aerosol size is more pronounced under weak forcing (wider spread in light blue shading). 

Likewise, $\beta_\text{ext}$ is insensitive to size under most regimes. Under strong forcing, $\beta_{\text{ext}}$ increases with aerosol loading and shows little variation with size even under highly polluted conditions. Under weakly forced conditions, on the other hand, the response in $\beta_{\text{ext}}$ to increased aerosol loading varies with aerosol size; whereas particles with a dry aerosol size of 50- to 100-nm lead to only small increases in $\beta_{\text{ext}}$ with $N$ under weak forcing, $\beta_{\text{ext}}$ shows larger increases with increasing $N$ for 250-nm particles. These results highlight that aerosol number primarily controls cloud microphysical and radiative properties, with aerosol size playing the secondary role depending on environmental and pollution regime.

\section*{Discussion and Conclusions}
While it has long been recognized that changes in aerosol loading can influence cloud droplet size distributions and reflectivity \citep{TwomeyandWarner, Twomey1974,Twomey1977,chandrakar2016aerosol,breon2002aerosol,nakajima2001possible,quaas2008satellite,efraim2020satellite}, our study provides new insights into the importance of aerosol characteristics in determining these perturbations to cloud properties. Using controlled laboratory experiments in the Pi Chamber and corresponding LES simulations, we show that cloud droplet size distributions and optical depths are often insensitive to the underlying aerosol size distribution. Instead, our results indicate that cloud droplet size distributions and optical properties are primarily controlled by the strength of convective forcing and the aerosol loading, which together determine the mean and variability of water vapor supersaturation. In the Pi Chamber, these correspond to the temperature gradient between the top and bottom plates and the aerosol injection rate, while in the atmosphere they more broadly map to meteorological updraft strength and pollution levels. This finding highlights that the sensitivity of cloud reflectivity to aerosol size is governed more by the environmental regime than by particle characteristics.

Our findings build on earlier work that has explored the impact of aerosol size and composition on CCN activity. Previous studies have shown that CCN activity depends strongly on particle size \citep{reutter2009aerosol,Dusek2006,Hudson2007,petters2007single,fierce2013cloud,fierce2015explaining}, and, under idealized cloud simulations, these differences in CCN activity translate to large differences in the number \citep{reutter2009aerosol}, size distribution \citep{Dusek2006,Hudson2007,chandrakar2016aerosol,hsieh2009parameterization}, and ultimately optical properties of the cloud droplet population \citep{grosvenor2018remote}. In contrast, we show that, under turbulent conditions that better represent natural clouds, size-induced differences in CCN activity often do not manifest in cloud microphysical or optical properties. Whereas previous studies concluded that CCN activity depends more strongly on size than composition \citep{Dusek2006}, we show that even size is frequently a poor predictor once turbulence and variability in supersaturation are taken into account.  

We found that the relative change in cloud optical properties to changes in aerosol size, quantified as $d\ln\beta{\text{ext}}/d\ln D_{\text{a}}$, is most pronounced under conditions of weak convective forcing and high aerosol loading. These conditions were characterized by relatively low mean supersaturation levels and weak turbulent fluctuations. Decreasing the aerosol loading within the chamber led to an increase in supersaturation levels, stronger fluctuations, and weaker response in the cloud droplet size distribution to the underlying aerosol size distribution. Under these conditions of low aerosol loading, even with weak forcing, $d\ln\beta_{\text{ext}}/d\ln D_{\text{a}}$ decreased to nearly 0, indicating that $\beta_{\text{ext}}$ is insensitive to aerosol size. The mean and variability in supersaturation was greatest --- and the sensitivity in the observed and modeled size distribution was weakest --- under conditions of strong forcing and low aerosol loading, suggesting that cloud droplet formation is not strongly influenced by solute effects of aerosol particles.

The key findings of our study are as follows:
\begin{enumerate}
    \item Cloud droplet size distributions are often insensitive to aerosol size, as shown consistently in Pi Chamber experiments and LES simulations;
    \item Cloud optical proprieties are generally unaffected by aerosol size, except under highly polluted and weakly forced conditions;
    \item Under strong forcing and low aerosol loading, high supersaturation levels suppress sensitivity to aerosol size. Conversely, under weak forcing and polluted conditions, supersaturation fluctuations dominate cloud formation, revealing conditions where aerosol size effects may be observed. 
\end{enumerate}

Together, these results suggest that in the real atmosphere, regions with high aerosol loading, such as urban and industrial areas, may experience more pronounced shifts in cloud properties with changes in aerosol size than cleaner environments. The dependence of sensitivity on turbulent forcing conditions highlights the importance of considering dynamic variability when evaluating aerosol–cloud interactions. By combining laboratory and numerical experiments, this work provides a comprehensive framework for understanding how aerosol variability shapes cloud formation and evolution, advancing our ability to assess human impacts on the Earth system. 

\section*{Methods}\label{sec:Methods} 

\subsection*{Experimental Setup}
To quantify the relationship between aerosol particle properties and cloud droplet size distributions in a controlled setting, we conducted experiments using the Pi Chamber, a convective chamber at Michigan Tech designed to study aerosol-cloud interactions in turbulent environments through moist Rayleigh-Bénard convection \cite{chandrakar2016aerosol,chang2016laboratory}.
For strong forcing cases, we set the temperature gradient between the bottom and top boundaries to $\Delta${T} = 18K. The gradient for weak forcing was $\Delta${T} = 9K. The surfaces of the chamber are held at liquid water saturation through the use of wet filter paper. To form a cloud, dry NaCl aerosol (diameters of 50~nm, 100~nm, and 250~nm ) were selected using an Aerodynamic Aerosol Classifier (AAC) and injected into the Pi Chamber at a constant flow rate of 5~lpm, with controlled concentrations of 10,000 cm$^{-3}$ and 100,000 cm$^{-3}$ for clean and polluted conditions respectively. (Note that the quoted concentrations are for the injected air stream.) After 30 minutes of injection, cloud conditions in the chamber reached a steady state, and the cloud droplet size distributions were measured using a Welas Digital 2000, an optical light-scattering spectrometer system, covering a size range of 0.6~$\mu${m} to 40~$\mu${m}. (See Supplementary Information for full Pi Chamber description.)

\subsection*{Model description}
We used the latest microphysics scheme version of the cloud convective chamber Large-Eddy Simulation (LES) model -System for Atmospheric Modeling (SAM)- which integrates the haze-capable bin microphysics scheme developed by Yang et al. \cite{yang2023intercomparison}. This scheme operates under the principle that all dry aerosol particles injected into the chamber during simulation initially become haze particles through deliquescence processes. Subsequently, they undergo activation to form cloud droplets, providing that the environmental supersaturation exceeds their critical supersaturation. The growth of these haze particles is governed by condensation, incorporating both curvature and solute effects. The distribution of the hydrometeors (including haze and cloud droplets) larger than 0.1 $\mu m$ in radius is represented by a total of 40 mass-doubling bins. 

The simulation setup closely follows of that of Yang et. al.\cite{yang2023intercomparison}, except for using a coarser grid spacing. Specially, the domain dimensions were set to $L_x=L_y=2$~m, and $L_z = 1$~m, with the grid resolution of $6.25$~cm and a time step of $0.02 s$ for a 1-hour simulation. This grid spacing falls within the inertial subrange of a cloud chamber \cite{wang2024investigation}. 
We run four different scenarios: (1) In the first two cases of simulation, the goal is to generate high mean supersaturation in the Pi Chamber as guided by chamber observations \cite{chandrakar2016aerosol}. This is achieved by setting the bottom temperature to T$_{bot} = 299$~K, the top temperature to T$_{top} = 280$~K, the sidewalls temperature equal to the mean temperature T$_{sidewalls} = 285 K$, with a temperature gradient of $\Delta T = 19$ K. Regarding the aerosol properties, three different monodisperse sodium chloride (NaCl) aerosols with the initial aerosol number concentrations of 100 mg$^{-1}$ and with diameters of 50 nm, 100 nm, and 250 nm are selected and injected into each grid box of the simulation domain at each time step. Two different aerosol injection rates of 0.25 mg$^{-1}$ s$^{-1}$ and 5.0 mg$^{-1}$ s$^{-1}$ are used to represent clean and polluted conditions, respectively. (2) In the third and fourth cases, we created a low mean supersaturation environment by decreasing the bottom temperature to T$_{bot}$ = 289 K, changing the sidewalls temperature at T$_{sidewalls}$ = 284.5 K, while other parameters remain the same as shown in the table ~\ref{tab:table1}.

\setlength{\tabcolsep}{8pt}
\renewcommand{\arraystretch}{1.3}
\begin{table}
    \centering
    \begin{tabular}{l c c c}
    \toprule
        Case Run & $\Delta T$ (in K) & Injection Rates (in mg$^{-1}$ s$^{-1}$) & $D_a$ (in nm)\\
    \midrule
         Strong Forcing, Clean     & 19 & 0.25 & 50, 100, 250 \\
         Weak Forcing, Clean       & 9  & 0.25 & 50, 100, 250 \\
         Strong Forcing, Polluted  & 19 & 5.0  & 50, 100, 250 \\
         Weak Forcing, Polluted    & 9  & 5.0 & 50, 100, 250 \\
    \bottomrule
    \end{tabular}
    \caption{Different scenarios used in the document: $\Delta T$ represents the temperature gradient,  $D_a$ denotes the dry diameter of the monodisperse injected aerosol of sodium chloride (NaCl) and the Injection Rates indicates of the source of dry aerosols}
    \label{tab:table1}
\end{table}

\subsection*{Calculation of Cloud Optical Depth}
To calculate the cloud optical proprieties (extinction coefficient $\beta_\text{ext}$), we used cloud droplet size distributions from both the experimental and simulations data. First, we computed the single-particle optical properties using the Python Mie scattering package (PyMieScatt). The input parameters included the wavelength of the incident visible light ($550~nm$), the complex refractive index of water droplets, and the cloud droplet diameter ($D_\text{p}$). This calculation provided the Mie extinction coefficient ($Q_\text{ext}$) for each droplet diameter, which we used to determine the extinction cross-section :
\begin{equation}\label{equation 1}
    C_\text{ext} = Q_\text{ext} .\pi (D_\text{p}/2)^2 
\end{equation}

Then we calculated the volume extinction coefficient by integrating ~\ref{equation 1} over the cloud droplet size distribution:
\begin{equation}\label{equation2}
    \beta_\text{ext} = \int C_\text{ext}.\frac{dN}{d\log D_\text{p}}d\log D_\text{p}
\end{equation}


To assess the sensitivity of cloud properties to changes in aerosol number concentrations, which can influence radiative forcing and climate, we quantified the relative change in $\beta_\text{ext}$ to the relative change in dry injected aerosol diameter to assess the sensitivity of $\beta_\text{ext}$ to variations of the pollution lever and the mixing forcing in the Pi Chamber as :
\begin{equation}\label{equation4}
    \frac{d\beta_{\text{ext}}/\beta_\text{ext}}{dD_{\text{a}}/D_{\text{a}}} \simeq {d\ln\beta_\text{ext}}/{d\ln D_{\text{a}}},
\end{equation}

where $D_a$ represents the dry injected aerosol diameter. 

\newpage
\section*{Acknowledgments}
K.F. Gali, R. Shaw, J.C. Anderson and W. Cantrell were supported through funding from the U.S. Department of Energy's Atmospheric System Research (DE-SC0022128). H.F. Sadi was supported through  National Science Foundation grant AGS-2133229. L. Fierce and A. Wang were supported by the U.S. Department of Energy's Atmospheric System Research, an Office of Science Biological and Environmental Research program. P. Beeler was supported through the Laboratory Directed Research and Development program (Linus Pauling Distinguished Postdoctoral Fellowship Program) of Pacific Northwest National Laboratory (PNNL). The Pacific Northwest National Laboratory (PNNL) is operated for DOE by Battelle Memorial Institute under contract DE-AC05-76RL01830. F.Yang was supported by Office of Science Biological and Environmental Research program as part of the Department of Energy Atmospheric Systems Research program under contract DE-SC0012704. D. Richter was supported by NSF grant AGS-2227012 and ONR grant N00014-21-1-2296. The High-Performance Computing Shared Facility (Superior) at Michigan Technological University was used to perform the high-resolution LES presented in this publication.

\section*{Author contributions}
K. F. Gali performed experiments and simulations, conducted analysis, contributed to the code, and wrote the paper. H.F. Sadi and J.C. Anderson performed the experiments and contributed to the data analysis. P. Beeler contributed to the code and analysis. A. Wang and D. Richter contributed to analysis. F. Yang provided the LES microphysics code, contributed to the data analysis, and contributed to writing the paper. R. Shaw conceived the experiment, contributed to the data analysis, and contributed to writing the paper. W. Cantrell and L. Fierce conceived the experiment and analysis, conducted analysis, contributed the code and wrote the paper. All authors discussed and reviewed the paper.

\section*{Correspondent author}
Correspondence should be addressed to Will Cantrell (\url{cantrell@mtu.edu}) and Laura Fierce (\url{laura.fierce@pnnl.gov})

\newpage
\bibliographystyle{unsrtnat}
\bibliography{References}

@article{yang2023intercomparison,
  title={An intercomparison of large-eddy simulations of a convection cloud chamber using haze-capable bin and {L}agrangian cloud microphysics schemes},
  author={Yang, Fan and Hoffmann, Fabian and Shaw, Raymond A and Ovchinnikov, Mikhail and Vogelmann, Andrew M},
  journal={Journal of Advances in Modeling Earth Systems},
  volume={15},
  number={5},
  pages={e2022MS003270},
  year={2023},
  publisher={Wiley Online Library}
}

@article{anderson2023enhancements,
  title={Enhancements in cloud condensation nuclei activity from turbulent fluctuations in supersaturation},
  author={Anderson, Jesse C and Beeler, Payton and Ovchinnikov, Mikhail and Cantrell, Will and Krueger, Steven and Shaw, Raymond A and Yang, Fan and Fierce, Laura},
  journal={Geophysical Research Letters},
  volume={50},
  number={17},
  pages={e2022GL102635},
  year={2023},
  publisher={Wiley Online Library}
}

@article{chang2016laboratory,
  title={A laboratory facility to study gas--aerosol--cloud interactions in a turbulent environment: The $\Pi$ chamber},
  author={Chang, K and Bench, J and Brege, M and Cantrell, Will and Chandrakar, K and Ciochetto, David and Mazzoleni, Claudio and Mazzoleni, LR and Niedermeier, Dennis and Shaw, RA},
  journal={Bulletin of the American Meteorological Society},
  volume={97},
  number={12},
  pages={2343--2358},
  year={2016}
}

@article{prabhakaran2020role,
  title={The role of turbulent fluctuations in aerosol activation and cloud formation},
  author={Prabhakaran, Prasanth and Shawon, Abu Sayeed Md and Kinney, Gregory and Thomas, Subin and Cantrell, Will and Shaw, Raymond A},
  journal={Proceedings of the National Academy of Sciences},
  volume={117},
  number={29},
  pages={16831--16838},
  year={2020},
  publisher={National Acad Sciences}
}

@article{thomas2019scaling,
  title={Scaling of an atmospheric model to simulate turbulence and cloud microphysics in the {P}i Chamber},
  author={Thomas, Subin and Ovchinnikov, Mikhail and Yang, Fan and van der Voort, Dennis and Cantrell, Will and Krueger, Steven K and Shaw, Raymond A},
  journal={Journal of Advances in Modeling Earth Systems},
  volume={11},
  number={7},
  pages={1981--1994},
  year={2019},
  publisher={Wiley Online Library}
}

@article{chandrakar2016aerosol,
  title={Aerosol indirect effect from turbulence-induced broadening of cloud-droplet size distributions},
  author={Chandrakar, Kamal Kant and Cantrell, Will and Chang, Kelken and Ciochetto, David and Niedermeier, Dennis and Ovchinnikov, Mikhail and Shaw, Raymond A and Yang, Fan},
  journal={Proceedings of the National Academy of Sciences},
  volume={113},
  number={50},
  pages={14243--14248},
  year={2016},
  publisher={National Acad Sciences}
}

@article{TwomeyandWarner,
  title={Comparison of measurements of cloud droplets and cloud nuclei},
  author={Twomey, S and Warner, J},
  journal={Journal of Atmospheric Sciences},
  volume={24},
  number={6},
  pages={702--703},
  year={1967}
}

@article{Twomey1974,
  title={Pollution and the planetary albedo},
  author={Twomey, Sean},
  journal={Atmospheric Environment (1967)},
  volume={8},
  number={12},
  pages={1251--1256},
  year={1974},
  publisher={Elsevier}
}

@article{Twomey1977,
  title={The influence of pollution on the shortwave albedo of clouds},
  author={Twomey, Sean},
  journal={Journal of the Atmospheric Sciences},
  volume={34},
  number={7},
  pages={1149--1152},
  year={1977},
  publisher={American Meteorological Society}
}

@article{Albrecht1989,
  title={Aerosols, cloud microphysics, and fractional cloudiness},
  author={Albrecht, Bruce A},
  journal={Science},
  volume={245},
  number={4923},
  pages={1227--1230},
  year={1989},
  publisher={American Association for the Advancement of Science}
}

@article{PincusandBaker,
  title={Effect of precipitation on the albedo susceptibility of clouds in the marine boundary layer},
  author={Pincus, Robert and Baker, Marcia B},
  journal={Nature},
  volume={372},
  number={6503},
  pages={250--252},
  year={1994},
  publisher={Nature Publishing Group UK London}
}

@article{Dusek2006,
  title={Size matters more than chemistry for cloud-nucleating ability of aerosol particles},
  author={Dusek, Ulrike and Frank, GP and Hildebrandt, Lea and Curtius, Joachim and Schneider, Johannes and Walter, Saskia and Chand, Duli and Drewnick, Frank and Hings, Silke and Jung, D and others},
  journal={Science},
  volume={312},
  number={5778},
  pages={1375--1378},
  year={2006},
  publisher={American Association for the Advancement of Science}
}

@article{Hudson2007,
  title={Variability of the relationship between particle size and cloud-nucleating ability},
  author={Hudson, James G},
  journal={Geophysical Research Letters},
  volume={34},
  number={8},
  year={2007},
  publisher={Wiley Online Library}
}

@article{Li2024GRL,
  title={On the prediction of aerosol-cloud interactions within a data-driven framework},
  author={Li, Xiang-Yu and Wang, Hailong and Chakraborty, TC and Sorooshian, Armin and Ziemba, Luke D and Voigt, Christiane and Thornhill, Kenneth Lee and Yuan, Emma},
  journal={Geophysical Research Letters},
  volume={51},
  number={24},
  pages={e2024GL110757},
  year={2024},
  publisher={Wiley Online Library}
}

@article{reutter2009aerosol,
  title={Aerosol-and updraft-limited regimes of cloud droplet formation: influence of particle number, size and hygroscopicity on the activation of cloud condensation nuclei ({CCN})},
  author={Reutter, Philipp and Su, H and Trentmann, J and Simmel, Martin and Rose, Diana and Gunthe, SS and Wernli, H and Andreae, MO and P{\"o}schl, U},
  journal={Atmospheric Chemistry and Physics},
  volume={9},
  number={18},
  pages={7067--7080},
  year={2009},
  publisher={Copernicus Publications G{\"o}ttingen, Germany}
}

@article{masson2021ipcc,
  title={{IPCC}, 2021: Summary for policymakers. in: Climate change 2021: The physical science basis. {C}ontribution of working group {I} to the sixth assessment report of the {I}ntergovernmental {P}anel on {C}limate {C}hange},
  author={Masson-Delmotte, VP and Zhai, Panmao and Pirani, SL and Connors, C and P{\'e}an, S and Berger, N and Caud, Y and Chen, L and Goldfarb, MI and Scheel Monteiro, Pedro M},
  year={2021},
  publisher={Cambridge University Press, Cambridge, United Kingdom and New York, NY, USA}
}

@article{lee2023ipcc,
  title={{IPCC}, 2023: Climate Change 2023: Synthesis Report, Summary for Policymakers. Contribution of Working Groups {I}, {II} and {III} to the Sixth Assessment Report of the Intergovernmental Panel on Climate Change [Core Writing Team, H. Lee and J. Romero (eds.)]. {IPCC}, Geneva, Switzerland.},
  author={Lee, Hoesung and Calvin, Katherine and Dasgupta, Dipak and Krinner, Gerhard and Mukherji, Aditi and Thorne, Peter and Trisos, Christopher and Romero, Jos{\'e} and Aldunce, Paulina and Barret, Ko and others},
  year={2023},
  publisher={Intergovernmental Panel on Climate Change (IPCC)}
}

@article{change2001working,
  title={Working Group {I} : The Scientific Basis},
  author={Change, Climate},
  journal={Third Assessment Report of the Intergovernmental Panel on Climate Change},
  year={2001},
  publisher={Cambridge University New York, NY, USA}
}

@article{portner2022ipcc,
  title={{IPCC}, 2022: Summary for policymakers},
  author={P{\"o}rtner, Hans-Otto and Roberts, Debra C and Poloczanska, Elvira S and Mintenbeck, Katja and Tignor, M and Alegr{\'\i}a, A and Craig, Marlies and Langsdorf, Stefanie and L{\"o}schke, Sina and M{\"o}ller, Vincent and others},
  year={2022},
  publisher={Cambridge University Pres}
}

@article{anderson2021AMT,
  title={Effects of the large-scale circulation on temperature and water vapor distributions in the $\Pi$ Chamber},
  author={Anderson, Jesse C and Thomas, Subin and Prabhakaran, Prasanth and Shaw, Raymond A and Cantrell, Will},
  journal={Atmospheric Measurement Techniques},
  volume={14},
  number={8},
  pages={5473--5485},
  year={2021},
  publisher={Copernicus GmbH}
}

@article{anderson2024JAS,
  title={Droplet growth or evaporation does not buffer the variability in supersaturation in clean clouds},
  author={Anderson, Jesse C and Helman, Ian and Shaw, Raymond A and Cantrell, Will},
  journal={Journal of the Atmospheric Sciences},
  volume={81},
  number={1},
  pages={225--233},
  year={2024},
  publisher={American Meteorological Society}
}

@article{petters2007single,
  title={A single parameter representation of hygroscopic growth and cloud condensation nucleus activity},
  author={Petters, MD and Kreidenweis, SM},
  journal={Atmospheric Chemistry and Physics},
  volume={7},
  number={8},
  pages={1961--1971},
  year={2007},
  publisher={Copernicus Publications G{\"o}ttingen, Germany}
}

@article{fierce2013cloud,
  title={When is cloud condensation nuclei activity sensitive to particle characteristics at emission?},
  author={Fierce, Laura and Riemer, Nicole and Bond, Tami C},
  journal={Journal of Geophysical Research: Atmospheres},
  volume={118},
  number={24},
  pages={13--476},
  year={2013},
  publisher={Wiley Online Library}
}

@article{fierce2015explaining,
  title={Explaining variance in black carbon's aging timescale},
  author={Fierce, Laura and Riemer, Nicole and Bond, Tami C},
  journal={Atmospheric Chemistry and Physics},
  volume={15},
  number={6},
  pages={3173--3191},
  year={2015},
  publisher={Copernicus GmbH}
}

@article{yang2022large,
  title={Large-eddy simulations of a convection cloud chamber: Sensitivity to bin microphysics and advection},
  author={Yang, Fan and Ovchinnikov, Mikhail and Thomas, Subin and Khain, Alexander and McGraw, Robert and Shaw, Raymond A and Vogelmann, Andrew M},
  journal={Journal of Advances in Modeling Earth Systems},
  volume={14},
  number={5},
  pages={e2021MS002895},
  year={2022},
  publisher={Wiley Online Library}
}

@article{wang2024designing,
  title={Designing a convection-cloud chamber for collision-coalescence using large-eddy simulation with bin microphysics},
  author={Wang, Aaron and Ovchinnikov, Mikhail and Yang, Fan and Schmalfuss, Silvio and Shaw, Raymond A},
  journal={Journal of Advances in Modeling Earth Systems},
  volume={16},
  number={1},
  pages={e2023MS003734},
  year={2024},
  publisher={Wiley Online Library}
}

@article{wang2024dual,
  title={The Dual Nature of Entrainment-Mixing Signatures Revealed Through Large-Eddy Simulations of a Convection-Cloud Chamber},
  author={Wang, Aaron and Ovchinnikov, Mikhail and Yang, Fan and Cantrell, Will and Yeom, Jaemin and Shaw, Raymond A},
  journal={Journal of the Atmospheric Sciences},
  volume={81},
  number={12},
  pages={2017--2039},
  year={2024},
  publisher={American Meteorological Society}
}

@article{wang2024glaciation,
  title={Glaciation of mixed-phase clouds: insights from bulk model and bin-microphysics large-eddy simulation informed by laboratory experiment},
  author={Wang, Aaron and Krueger, Steve and Chen, Sisi and Ovchinnikov, Mikhail and Cantrell, Will and Shaw, Raymond A},
  journal={Atmospheric Chemistry and Physics},
  volume={24},
  number={18},
  pages={10245--10260},
  year={2024},
  publisher={Copernicus Publications G{\"o}ttingen, Germany}
}

@article{yang2025microphysics,
  title={Microphysics regimes due to haze--cloud interactions: cloud oscillation and cloud collapse},
  author={Yang, Fan and Sadi, Hamed Fahandezh and Shaw, Raymond A and Hoffmann, Fabian and Hou, Pei and Wang, Aaron and Ovchinnikov, Mikhail},
  journal={Atmospheric Chemistry and Physics},
  volume={25},
  number={6},
  pages={3785--3806},
  year={2025},
  publisher={Copernicus Publications G{\"o}ttingen, Germany}
}

@article{breon2002aerosol,
  title={Aerosol effect on cloud droplet size monitored from satellite},
  author={Br{\'e}on, Francois-Marie and Tanr{\'e}, Didier and Generoso, Sylvia},
  journal={Science},
  volume={295},
  number={5556},
  pages={834--838},
  year={2002},
  publisher={American Association for the Advancement of Science}
}

@article{hsieh2009parameterization,
  title={Parameterization of cloud droplet size distributions: Comparison with parcel models and observations},
  author={Hsieh, Wei-Chun and Nenes, Athanasios and Flagan, Richard C and Seinfeld, John H and Buzorius, G and Jonsson, H},
  journal={Journal of Geophysical Research: Atmospheres},
  volume={114},
  number={D11},
  year={2009},
  publisher={Wiley Online Library}
}

@article{grosvenor2018remote,
  title={Remote sensing of droplet number concentration in warm clouds: A review of the current state of knowledge and perspectives},
  author={Grosvenor, Daniel P and Sourdeval, Odran and Zuidema, Paquita and Ackerman, Andrew and Alexandrov, Mikhail D and Bennartz, Ralf and Boers, Reinout and Cairns, Brian and Chiu, J Christine and Christensen, Matthew and others},
  journal={Reviews of Geophysics},
  volume={56},
  number={2},
  pages={409--453},
  year={2018},
  publisher={Wiley Online Library}
}

@article{nakajima2001possible,
  title={A possible correlation between satellite-derived cloud and aerosol microphysical parameters},
  author={Nakajima, Teruyuki and Higurashi, Akiko and Kawamoto, Kazuaki and Penner, Joyce E},
  journal={Geophysical Research Letters},
  volume={28},
  number={7},
  pages={1171--1174},
  year={2001},
  publisher={Wiley Online Library}
}

@article{quaas2008satellite,
  title={Satellite-based estimate of the direct and indirect aerosol climate forcing},
  author={Quaas, Johannes and Boucher, Olivier and Bellouin, Nicolas and Kinne, Stefan},
  journal={Journal of Geophysical Research: Atmospheres},
  volume={113},
  number={D5},
  year={2008},
  publisher={Wiley Online Library}
}

@article{efraim2020satellite,
  title={Satellite retrieval of cloud condensation nuclei concentrations in marine stratocumulus by using clouds as {CCN} chambers},
  author={Efraim, Avichay and Rosenfeld, Daniel and Schmale, Julia and Zhu, Yannian},
  journal={Journal of Geophysical Research: Atmospheres},
  volume={125},
  number={16},
  pages={e2020JD032409},
  year={2020},
  publisher={Wiley Online Library}
}

@article{wang2024investigation,
  title={An investigation of {LES} wall modeling for {R}ayleigh--{B}{\'e}nard convection via interpretable and physics-aware feedforward neural networks with DNS},
  author={Wang, Aaron and Yang, Xiang IA and Ovchinnikov, Mikhail},
  journal={Journal of the Atmospheric Sciences},
  volume={81},
  number={2},
  pages={435--458},
  year={2024},
  publisher={American Meteorological Society}
}

@article{hudson2020ccn,
  title={CCN spectral shape and cumulus cloud and drizzle microphysics},
  author={Hudson, James G and Noble, Stephen},
  journal={Journal of Geophysical Research: Atmospheres},
  volume={125},
  number={1},
  pages={e2019JD031141},
  year={2020},
  publisher={Wiley Online Library}
}

@article{che2017prediction,
  title={Prediction of size-resolved number concentration of cloud condensation nuclei and long-term measurements of their activation characteristics},
  author={Che, HC and Zhang, XY and Zhang, L and Wang, YQ and Zhang, YM and Shen, XJ and Ma, QL and Sun, JY and Zhong, JT},
  journal={Scientific reports},
  volume={7},
  number={1},
  pages={5819},
  year={2017},
  publisher={Nature Publishing Group UK London}
}

@article{allwayin2024locally,
  title={Locally narrow droplet size distributions are ubiquitous in stratocumulus clouds},
  author={Allwayin, Nithin and Larsen, Michael L and Glienke, Susanne and Shaw, Raymond A},
  journal={Science},
  volume={384},
  number={6695},
  pages={528--532},
  year={2024},
  publisher={American Association for the Advancement of Science}
}

@article{fanourgakis2019evaluation,
  title={Evaluation of global simulations of aerosol particle and cloud condensation nuclei number, with implications for cloud droplet formation},
  author={Fanourgakis, George S and Kanakidou, Maria and Nenes, Athanasios and Bauer, Susanne E and Bergman, Tommi and Carslaw, Ken S and Grini, Alf and Hamilton, Douglas S and Johnson, Jill S and Karydis, Vlassis A and others},
  journal={Atmospheric chemistry and physics},
  volume={19},
  number={13},
  pages={8591--8617},
  year={2019},
  publisher={Copernicus GmbH}
}

@article{tang2023earth,
  title={Earth System Model Aerosol--Cloud Diagnostics (ESMAC Diags) package, version 2: assessing aerosols, clouds, and aerosol--cloud interactions via field campaign and long-term observations},
  author={Tang, Shuaiqi and Varble, Adam C and Fast, Jerome D and Zhang, Kai and Wu, Peng and Dong, Xiquan and Mei, Fan and Pekour, Mikhail and Hardin, Joseph C and Ma, Po-Lun},
  journal={Geoscientific Model Development},
  volume={16},
  number={21},
  pages={6355--6376},
  year={2023},
  publisher={Copernicus GmbH}
}

\end{document}